\documentstyle[aps,prl,multicol,epsf,epsfig]{revtex}

\begin{document}
\def\CC{{\rm\kern.24em \vrule width.04em height1.46ex depth-.07ex
\kern-.30em C}}
\def\P{{\rm I\kern-.25em P}}
\def\RR{{\rm
         \vrule width.04em height1.58ex depth-.0ex
         \kern-.04em R}}

\newcommand{\be}{\begin{equation}}
\newcommand{\ee}{\end{equation}}
\newcommand{\bq}{\begin{eqnarray}}
\newcommand{\eq}{\end{eqnarray}}
\newcommand{\Sc}{Schr\"odinger\,\,}
\newcommand{\Sp}{\,\,\,\,\,\,\,\,\,\,\,\,\,}
\newcommand{\no}{\nonumber\\}
\newcommand{\tr}{\text{tr}}
\newcommand{\p}{\partial}
\newcommand{\la}{\lambda}
\newcommand{\La}{\Lambda}
\newcommand{\G}{{\cal G}}
\newcommand{\D}{{\cal D}}
\newcommand{\E}{{\cal E}}
\newcommand{\W}{{\bf W}}
\newcommand{\de}{\delta}
\newcommand{\al}{\alpha}
\newcommand{\bi}{\beta}
\newcommand{\ep}{\varepsilon}
\newcommand{\ga}{\gamma}
\newcommand{\epp}{\epsilon}
\newcommand{\vep}{\varepsilon}
\newcommand{\th}{\theta}
\newcommand{\om}{\omega}
\newcommand{\si}{\sigma}
\newcommand{\J}{{\cal J}}
\newcommand{\pr}{\prime}
\newcommand{\ka}{\kappa}
\newcommand{\TH}{\mbox{\boldmath${\theta}$}}
\newcommand{\DE}{\mbox{\boldmath${\delta}$}}
\newcommand{\lan}{\langle}
\newcommand{\ran}{\rangle}
\newcommand{\Hol}{\text{Hol}}
\newcommand{\cp}{{\bf CP}}
\newcommand{\1}{{\bf 1}}
\newcommand{\U}{{\cal U}}


%
\newcommand{\bra}[1]{\left<#1\right|} \newcommand{\ket}[1]{\left|#1\right>}
\newcommand{\braket}[1]{\left<#1\right>}
\newcommand{\inner}[2]{\left<#1|#2\right>}
\newcommand{\sand}[3]{\left<#1|#2|#3\right>}
\newcommand{\proj}[2]{\left|#1\left>\right<#2\right|} %
\newcommand{\rbra}[1]{\left(#1\right|} \newcommand{\rket}[1]{\left|#1\right)}
\newcommand{\rbraket}[1]{\left(#1\right)}
\newcommand{\rinner}[2]{\left(#1|#2\right)}
\newcommand{\rsand}[3]{\left(#1|#2|#3\right)}
\newcommand{\rproj}[2]{\left|#1\left)\right(#2\right|}
\newcommand{\absqr}[1]{{\left|#1\right|}^2}
\newcommand{\abs}[1]{\left|#1\right|}
\newcommand{\plz}[1]{\partial_{z}^{#1}}
\newcommand{\plzb}[1]{\partial_{\overline{z}}^{#1}} 
\newcommand{\zi}[1]{ z^{#1}}
\newcommand{\zib}[1]{{\overline{z}}^{#1}} 
\newcommand{\mat}[4]{\left(\begin{array}{cc} #1 & #2 \\ #3 & #4
\end{array}\right)} %
\newcommand{\col}[2]{\left( \begin{array}{c} #1 \\ #2
\end{array} \right)} 
\def\a{\alpha} 
\def\b{\beta} 
\def\g{\gamma}
\def\d{\delta} 
\def\e{\epsilon} 
\def\z{\zeta} 
\def\th{\theta} 
\def\f{\phi}
\def\la{\lambda} 
\def\m{\mu} 
\def\om{\omega}  
\def\D{\Delta} 
\def\zb{\bar{z}} 
\newcommand{\lag}[2]{L_{#1}^{#2}(4\l^{2})}
\newcommand{\mes}[1]{d\mu(#1)} 
\def\nd{\noindent}
\def\non{\nonumber} 
\def\cap{\caption} 
\def\cline{\centerline}
\newcommand{\ba}{\begin{array}} 
\newcommand{\ea}{\end{array}}
\newcommand{\bea}{\begin{eqnarray}} 
\newcommand{\eea}{\end{eqnarray}}
\newcommand{\beann}{\begin{eqnarray*}} 
\newcommand{\eeann}{\end{eqnarray*}}
\newcommand{\bfg}{\begin{figure}} 
\newcommand{\efg}{\end{figure}}

\draft
\title{Universal Quantum Computation by 
Holonomic and Non-Local Gates with Imperfections}
\author{Demosthenes Ellinas$^{1}$
\footnote{Electronic address: ellinas@science.tuc.gr} and Jiannis Pachos$^{2,3}$ 
\footnote{Electronic address: jip@mpq.mpg.de }
}
\address{
$^{1}$ Technical Univeristy of Crete, Department of Sciences, Section of Mathematics
GR - 731 00 Chania, Crete Greece \\
$^{2}$ Max Planck Institut f\"ur Quantenoptik, D-85748 Garching, Germany\footnote{Present Address}\\
$^{3}$ Institute of Scientific Interchange Foundation, Villa Gualino, Viale Settimio Severo 65, I-10133 Torino, Italy\\
}
\date{\today}
\maketitle
\begin{abstract}
We present a non-local construction of universal gates by means of 
holonomic (geometric) quantum teleportation. 
The effect of the errors from an imperfect control of the
classical parameters, the looping variation of which builds up holonomic gates,
is investigated. Additionally, the influence of quantum decoherence on holonomic 
teleportation used as a computational primitive is studied. Advantages of the holonomic 
implementation with respect to control errors and dissipation are presented.

\end{abstract}


\begin{multicols}{2}
\narrowtext

Holonomic quantum computation $(HQC)$ is a mathematically  fascinating area where elements from 
differential geometry are employed in order to describe the logical
evolution of
a quantum system with (multiple) degenerate 
energy eigen-states \cite{ZARA,PAZA,ams,Erik}.
Recent works \cite{Ekert,PaChou,Bose} support the belief that holonomic
implementation in NMR, quantum optics or ion traps may be a possible avenue for quantum computation.
Here our aim is twofold: first, the scope of $HQC$ is extended to the
general framework of universal quantum computation
with the construction of non-local gates; second, non ideal $HQC$ implementations, in the form of
geometrical imperfections of adiabatic parametric closed paths needed for the
construction of holonomic gates, as well as the effect of the
simultaneous presence of decoherence, is formulated and investigated. Our choice for basing the whole construction on the 
error-avoiding paradigm of holonomic quantum wiring stems from the fact that
due to the geometric nature of making the irreducible connections
corresponding to holonomic gates, the latter are expected to be resilient to
certain types of errors. As this work shows, under conditions specified below this turns out to be
true. 

Specifically, the methodology for the generation of holonomies is as follows.
In the control parametric manifold of iso-spectral 
transformations of a given degenerate Hamiltonian closed paths (loops) are run
adiabatically in order to 
represent the evolution operation for each degenerate eigenspace as a holonomy
of a given connection, $A$ \cite{SHWI,WIZE,NAK}. The latter has a form defined from the 
structure of the bundle of the energy degenerate spaces \cite{PAZARA} and
its irreducibility manifests the universality of the holonomic gates.
The loops, when subject to imperfections
while spanned, introduce an error in the final gates through their accordingly 
fluctuated parameters. This is systematically studied for
the Hadamard (H) and the control-not (CN) gates, which are used to build up a
teleportation circuit robust to control errors.
Also the H and CN gates are constructed non locally by using
the teleportation circuit as a building primitive. The choice for 
implementing only these two gates is based on the following fact;
although H and CN belong in the Pauli group $C_1$ and the Clifford group $C_2$  
respectively, and need to be supplemented by an element of the class of gates
$C_3 \equiv \{{U \, / \, UC_{1} U^\dagger }\subseteq C_2 \}$ such as the $T$ Toffoli
gate, the $\pi/8$ gate (rotation about the $z$-axis by an angle $\pi/4$), or the 
controlled-phase gate ($\text{diag}(1,1,1,i)$), in order to achieve
universality \cite{gott,nielchua}, some of them are proven to be constructable
fault-tolerantly in circuits involving only measurements of $C_2$ gates (see e.g.
\cite{shor} for the $T$ gate, and \cite{Zhou} for 
similar construction of the other gates). In this way we can obtain universality
by studying the H and CN gates only.
As a figure of merit of the effect of both geometric imperfections and  
quantum dissipation on the overall performance of the teleportation circuit
its fidelity is investigated for various limiting values of the decoherence
and the imperfection parameters. 

As a starting point we shall use the $\cp^n$ holonomic model \cite{PAZARA} in order to acquire 
the desired quantum gates \cite{compl}. In \cite{PAZARA} a mathematical way for 
the holonomic construction of 
given gates by running specific loops in the control parametric manifold is presented. 
The initial Hamiltonian of this model is given by $H_0=\ep_0 | n+1\rangle \langle n+1 |$
which acts on the state-space spanned by $\{|\al \ran \}_{\alpha=1}^{n+1}$.
We assume that it is possible by external control to perform
equivalent transformations of $H_0$ given by
${\cal O}(H_0):=\{ {\cal U}\,H_0\,{\cal U}^\dagger\,/\,{\cal U}\in U(n+1)\}$. 
Their parametric space is isomorphic to the $n$-dimensional complex projective space 
${\bf{CP}}^n\cong U(n+1)/(U(n)\times U(1))$ which is at
the disposal of the experimenter.
Each point, ${\bf z}$, of the $2n$ dimensional ${\bf{CP}}^n$ manifold 
corresponds to a unitary matrix
${\cal U}({\bf z}) =U_1(z_1)U_2(z_2)...U_n(z_n)$, where
$U_\al(z_\al)=\exp [ G_\al(z_\al)]$ with 
$G_\al(z_\al)=z_\al |\al\ran\lan n + 1| -\bar z _\al |n ~+~ 1 \ran \lan \al |$ 
and $z_\al=\th_\al e^{i \phi_\al}$, for $\al=1,...,n$. 
If $|\psi\rangle_{in}$ is the initial state in the zero degenerate space of the Hamiltonian, 
at the end of the adiabatic run of a loop $C$ in the control manifold $\cp^n$ one
obtains $ |\psi\rangle _{out}= \Gamma_{A}(C) \,|\psi\rangle_{in}$.
The holonomy $\Gamma_{A}(C)\in U(n)$ has a geometric origin and its appearance
accounts for the non-trivial curvature of the bundle of eigenspaces 
over $\cp^n$.
By introducing the Wilczek-Zee connection \cite{WIZE}
$A^{\mu}_{\bar \alpha \al}:= \langle \bar \alpha|{\cal U}^\dagger({\bf z})
\,{\partial \over \partial{\bf z}_\mu}\,
{\cal U}({\bf z})|\al\rangle \,\, ,
$
with $\al, \bar \al =1,...,n$,
one finds
$\Gamma_{A}(C) ={\bf{P}}\exp \int_C A$ \cite{SHWI},
where ${\bf{P}}$ denotes path ordering.

For particular loops the following holonomies are calculated \cite{PAZARA};
for the loop $C_1 \in (\th_\bi, \phi_\bi)$, we obtain an abelian like holonomy,
$\Gamma_A(C_1)= e^{ -i  \Sigma_1 } |\bi \ran \lan \bi| +|\bi^\perp\ran \lan \bi^\perp |$,
where ${\cal H}= \text{span}\{|\bi \ran ,|\bi ^\perp \ran\}$. The area
$\Sigma_1=\int_{D(C_1)} d\th_{\beta} d \phi_{\beta} \cos \th_{\beta}$ may be represented as the one of the surface
enclosed by $C_1$ on a $S^2$ sphere with coordinates $(2\th_\bi,\phi_\bi)$, while $D(C_1)$ is
the enclosed surface on the $(\th_\bi,\phi_\bi)$ plane.
For $C_2 \in (\th_\bi, \phi_{\bar \bi})$, $\bar \bi > \bi$, we take
$\Gamma_A(C_2)= e^{ i  \Sigma_2 } |\bar \bi \ran \lan \bar \bi| +|\bar \bi^\perp\ran \lan \bar \bi^\perp |$
which is of similar abelian nature as $\Gamma_A(C_1)$.
In order to obtain a non-abelian holonomy we perform the loop
$C_3$ on the plane $(\th_\bi, \th_{\bar \bi})$ positioned at $\phi_\bi=\phi_{\bar\bi}=0$, 
resulting to
$\Gamma_A(C_3)= \exp [ -i(-i|\bi \ran \lan \bar \bi| +i |\bar \bi \ran \lan \bi |)\tilde \Sigma_3 ]$,
while by taking the plane $(\th_\bi, \th_{\bar \bi})$ to be at the position 
$\phi_\bi=\pi /2$ and $\phi_{\bar \bi}=0$, we obtain 
$\Gamma_A(C_4)= \exp [-i (| \bi \ran \lan \bar \bi| + |\bar \bi \ran \lan \bi |) \tilde \Sigma _4 ]$
where $\tilde \Sigma = \int_{D(C)} d\th_{\beta} d \theta_{\bar \beta} \cos \th_{\bar \beta}$.
The identity action on the rest of the states is implied. 

With these control manipulations holonomies are produced, which can be used as logical gates
with parameters the areas $\Sigma$.
In fact we obtain a whole set of closed paths in the parametric manifold
which give the same holonomies, as deformations of the loop shape and position give
the same gate provided their enclosed area is preserved.
It is worth noticing that the composition rules \cite{ZARA} of loops of multiplied holonomies
may reduce the total length of the transversed paths by combining loops on the
same or perpendicular planes for successive gates eventually reducing 
the required resources for the overall circuit. 

{\em Imperfect holonomies}. In order to study the errors introduced by imperfect 
control of the external parameters we adopt an imperfectly spanned loop, $C'$.
If the errors are statistical rather than systematical then the
area spanned by this loop are, to the first order, zero. 
Let us consider how systematic errors in the area effect one and two qubit gates.
The Hadamard gate is given by
$ \begin{array}{cc}
U_H=&  
 \left[  \begin{array}{ccc}    \cos \Sigma & \sin \Sigma \\
                \sin \Sigma & -\cos \Sigma \\
\end{array} \right] \,\, , 
\end{array}
$
for $\Sigma=\pi/4$. Up to a corrective phase given by $\Gamma_A(C_1)$ with $\Sigma_1=\pi$
it may be produced by a loop $C_3$,
with spanning area given by $\Sigma =\int_{D(C_3)} d\th_1 d \th_2 \cos \th_1$ 
with $D(C_3)$ taken to be a rectangular surface enclosed by 
$\{0 \leq \th_1 \leq \pi /2 , 0\leq \th_2 \leq \pi /4\}$.
Introduce an error in this surface by translating the boarders of $\th_1$ and
$\th_2$ by $\al$ and $\bi$ respectively, where $\al,\bi \ll 1$.
This is a kind of systematic error.
The imperfect Hadamard gate is given to the first order in $\ep$ by
$U_H (\ep)=U_H+\ep h$, with 
$\begin{array}{cc}
h=&{1 \over \sqrt{2}}  
 \left[  \begin{array}{ccc}    -1 & 1 \\
                 1 & 1 \\
\end{array} \right]
\end{array}$.
The integrand $\cos \th_1$ in $\Sigma$ is the $(\th_1,\th_2)$ dependent part
of the component of the field strength
$F_{\th_1 \th_2}=\p_{\th_1} A_{\th_2} -\p_{\th_2} A_{\th_1} +[A_{\th_1}, A_{\th_2}]$.
As another interpretation of the holonomy, within this approach, is the exponential 
of the flux of $F$ \cite{flux}, then we want this flux to be stable with respect to small 
deformations of the relevant surface. Hence, we can take this surface to be such that
fluctuations of its area give insignificant variations to the total flux.
Indeed, the flux enclosed by the deformed loop $C_3'$ is given by $\Sigma(\ep)
\approx {\pi \over 4} +\ep$, times the Pauli matrix $\sigma_2$,
with infinitesimal deviation $\ep =\bi$, where the infinitesimal $\al$ 
does not appear at all in the first order due to the choice of the position of
the rectangular's sides. 
Ideally we would like $F_{\th_1 \th_2}$ to be exponentially decreasing with respect to the
distance from a particular point of the control manifold  
so that for large loops centered at that point local deformations of the loop shape would not
alter the enclosed flux. A model with such characteristics may be
build with optical devices. Indeed, in \cite{PaChou} the one qubit
gates $\Gamma_A(C_I)=\exp -i\hat \sigma_1 \Sigma_I$ with $C_I \in
\left. (x,r_1)\right._{\th_1=0}$ and $\Sigma_I:=\int_{\Sigma(C_I)}
\!dxdr_1 2 e^{-2r_1}$ as well as $\Gamma_A(C_{II})=\exp -i\hat
\sigma_2 \Sigma_{II}$ with $C_{II} \in
\left. (y,r_1)\right._{\th_1=\pi}$ and
$\Sigma_{II}:=\int_{\Sigma(C_{II})} \! dydr_1 2 e^{-2r_1}$, which can
produce any one qubit operation, give for large values of the
squeezing parameter $r_1$ zero error in all orders of the loop
deformation along $r_1$. This can be considered as an initial point
for passing from geometrical QC to topological QC
\cite{Kitaev,Preskill}. Further study is needed for the construction of optical
(bosonic) two qubit gates with topological character. 

On the other hand, the control-not gate is given, up to phase corrections
again by a $C_3$ loop between the proper $(\th_\bi, \th_{\bar \bi})$ variables
with $\Sigma=\pi/2$. By inserting the errors
$(\al', \bi')$ in the corresponding components the area becomes $\Sigma(\de)\approx
{\pi \over 2} +\delta$, with $\de=\bi'\ll1$ the error deviation.
Hence, $U_{CN}(\de)=U_{CN}-i\de |1\ran \lan 1|\otimes {\bf 1}$.

\begin{figure}[ht]
  \epsffile{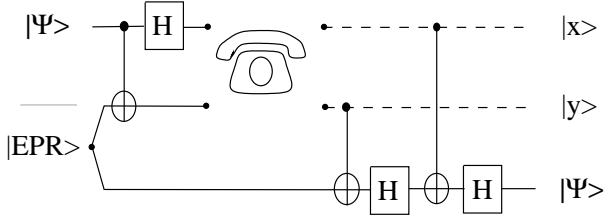}
  \caption[contour]{\label{tele}
The Brassard et.al. teleportation circuit. Dashed lines represent classical channels.
           }
\end{figure}

{\em Teleportation Circuit}. Consider the teleportation circuit \cite{Bra}
of Fig. \ref{tele}.
Depicted are the unknown state $|\Psi\ran= a_0|0\ran +a_1|1\ran$ which we wish to teleport,
the initially employed EPR states $|EPR\ran={1 \over \sqrt{2}} (|00\ran +|11\ran)$,
while the classical states $|x\ran$ or $|y\ran$ are the outcomes of measurement 
in the middle of the circuit.
For a more realistic implementation we may use the imperfect
Hadamard and CN gates as presented above, where we can take pertubatively into account
the imperfections of the Alice and
Bob holonomies in order to estimate
the reduction of the optimal fidelity of the above scheme,
due to imperfect holonomic implementation.
The overall circuit is represented as 
$\U_{tel}(\ep,\de)= \U_6(\ep)\U_5(\de) \U_4(\ep)\U_3(\de) \U_2(\ep) \U_1(\de)$,
where, $\ep$ is the error for the Hadamard gate and $\de$ is the error for the
CN gate. From the six gates presented in Fig. \ref{tele}, 
the first CN gate is written as 
\bq
{\cal U}_1(\de)&&= U_{CN}(\de) \otimes \1 \approx 
U_{CN}\otimes \1 -i \de |1\ran\lan1|\otimes \1 \otimes \1
\no \no
&&
\equiv \U_1+\de V_1
\nonumber
\eq
which is a unitary matrix up to order ${\cal O}(\ep)$.
The rest CN's are written similarly. 
The first Hadamard gate is given by
\bq
{\cal U}_2(\ep)&&= U_H(\ep) \otimes \1 \otimes \1\approx U_H \otimes \1 \otimes \1 +h \otimes \1 \otimes \1 
\no \no
&&
\equiv{\cal U}_2 +\ep V_2
\nonumber
\eq
and equivalently for $\U_4(\ep)$ and $\U_6(\ep)$.
The overall circuit, up to the first order in the error $\ep$ or $\de$ is given by
$\U_{tel}(\de,\ep)= \U_{tel} +\de(V_1+V_3+V_5)+\ep(V_2 + V_4+V_6)\equiv\U_{tel} +\de V_\de +\ep V_\ep$.
To quantify the error of the teleported state due to the imperfections in the loop spanning
we introduce the fidelity
\bq
&&
{\cal F}^{xy}_{\de, \ep} =\text{min}_{|\Psi\ran}
\left| \lan xy \Psi|\U_{tel} (\de,\ep)|\Psi EPR\ran \right|^2
\nonumber
\eq
where $|xy\ran$ is one of the possible outcomes $|00\ran$, $|01\ran$, 
$|10 \ran$ or $|11\ran$ for the two first qubits due to measurement.
In particular we find after minimization and tracing the different possibilities of 
$|xy \ran$ the result
\be
{\cal F}_{\de, \ep} =1 -\ep {3\over 2} 
(\sqrt{2} -1) -\de {1 \over 2 \sqrt{2}}
\label{fid}
\ee
which is smaller or equal to identity for small positive values of $\ep$ and $\de$.
Note that the coefficient of the CNs' error is almost {\it half} of the one of
the H gates, which is advantageous as for the two qubit gates you need
controllability of two qubit manipulations which should double
$\delta$ with respect to $\epsilon$.

{\em Application.} Let us proceed by adopting the teleportation as a kind of  
computation primitive, which can accept a proper input state (quantum software)
in a given site and provide output in another site a universal set of quantum gates,
in a fault-tolerant way. The teleportation method for gate construction is
adopted here as it essentially reduces the
needed resources for quantum computation to special ancilla state preparation
and also provides unifying 
technique for building in a fault-tolerant way a hierarchy of quantum gates
\cite{gott,Zhou,Nielsen}.

\begin{figure}[h]
  \epsffile{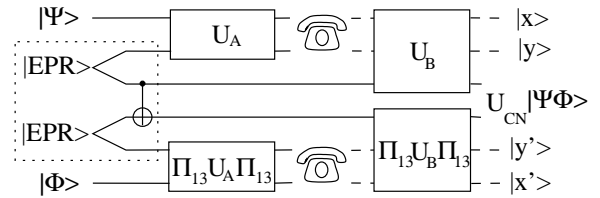}
  \caption[contour]{\label{cn}
The CN teleported gate. The dotted box includes the initial state sub-circuit,
           which can be implemented fault tolerantly.
           }
\end{figure}

Our aim is to use the Brassard's et.al. circuit of teleportation in order to produce 
(teleport) H and CN gates, which can be initially implemented on the EPR states 
fault-tolerantly. For that we consider imperfect holonomic
realization of these generalized circuits and  within first order
approximation, evaluate their robustness by obtaining their fidelity. 
%
Due to the similarity of the operators 
involved the derived fidelities are closely related with the one in (\ref{fid}).
Indeed, for the Hadamard gate the employed circuit is $U_{tel}^H= (\1 \otimes \1 \otimes 
U_H) \,U_{tel} \, (\1 \otimes \1 \otimes U_H^{\dagger}) $ and its fidelity 
${\cal F} _H$ with respect to the
transformed initial and final states, as can easily be shown, is equal to the fidelity of the
teleportation circuit itself, i.e ${\cal F} _H={\cal F}$.
For constructing the CN teleported gate we shall employ two
teleportation circuits with the additional permutation operator 
$\Pi_{13}=\sum_{x,y=0,1} |y\ran \lan x| \otimes \1 \otimes |x\ran \lan y|$.
The circuit of the two teleportations (see Fig. \ref{cn}) is 
arranged as $W_{tel}=U_{tel} \otimes
\Pi_{13} U_{tel} \Pi_{13}$ and the CN implementation is given by
$W_{tel} ^{CN} \equiv U_{CN}^{34} W_{tel} \left. U^{34}_{CN}\right.^{\dagger}$.
By calculating the fidelity of this circuit with respect
to the initial and final states we obtain ${\cal F}_{CN}(\ep ,\de)={\cal F}(2 \ep ,2 \de)$;
that is, it has the same functional form as in the case of one teleportation 
but the errors are now doubled. These results are valid to all orders
in $\epsilon$ and $\delta$. 

{\em Dissipation.} Let us now consider the case where imperfections in the
construction
of holonomic gates as have been studied so far are present
simultaneously with dissipative mechanisms in the modeling
of the circuits. As dissipation
is an almost unavoidable destruction of quantum coherence
that affects the performance quality of logical circuits,
it is expected to cooperate with the possible imperfections
in lowering their fidelity. To quantify these
thoughts we shall formulate the appearance of a general
class of dissipative mechanisms in the computational primitive
element of an imperfect teleportation circuit. More specifically
we shall study decoherence on the Bob's part of the total
density operator of the teleportation scheme, that takes place
after the completion of Alice's part of the circuit and during the
time she classically transmits two bits of information to Bob.
This can be also thought of as an imperfection during the measurement
procedure in the middle of the circuit.
Let $\rho_I = \proj{\Psi}{\Psi}\otimes\rho$, the initial density
operator with $\ket{\Psi} = \a_0 \ket{0}+\a_1\ket{1}$ the state
to be transmitted, and $\rho$ the transmitting density operator,
which in general can be taken not to be a perfect projector of
an $EPR$ entangled pair. Let $\overline{\cal H}={\cal H}\otimes
{\cal H}\otimes{\cal H}$, the total Hilbert space of Alice and Bob,
where ${\cal H}=span\left\{\ket{0},\ket{1} \right\}$.
Then let ${\cal P}=
End(\overline{\cal H})$, the space of pure density operators acting
on $\overline{\cal H}$, and ${\cal S}=hull({\cal P})$, the convex hull
of $\cal P$. Consider the linear, trace preserving and
completely positive family of maps  $\left \{ s_\lambda :{\cal S}\rightarrow {\cal S}
: \lambda \geq 0 \right \}$,
that admits a Kraus operator-sum representation such that
$s_\lambda (\rho)=\sum_{i=1}^{k}W_i \rho W_i^\dagger $, where
$\left\{W_i \right\}_{i=1}^{k} \in End(\overline{\cal H})$ and
$\sum_{i=1}^{k} W_i^\dagger W_i = {\bf 1}$. Since we are interested
in dissipation occurring in Bob's site only, we take $W_i ={\bf 1}
\otimes{\bf 1}\otimes V_i $, for some chosen $V_i$'s. 
Moreover,
the dissipation generators $W_i (\lambda)$ may be taken to depend on
the parameter $\lambda$, in such a way that
$\lim_{\lambda \rightarrow 0}W_1 (\lambda)={\bf 1}$,
$\lim_{\lambda \rightarrow 0}W_i (\lambda)={\bf 0}$, $i\neq 1$,
namely in the zero dissipation limit $s_{\lambda=0} (\rho)=\rho$.
Then we
rewrite $s_\lambda (\rho)=\sum_{i=1}^{k}Ad(W_i )\rho $, where
the adjoint action $Ad(X)\rho \equiv X\rho X^\dagger $ is employed.
By means of the property $Ad(XY)=Ad(X)Ad(Y)$, we now introduce
a $3$-parameter POVM $\left \{ {\cal \mu}_{\d , \ep , \lambda}:
{\cal S}\rightarrow{\cal S} : 0 \leq \d \leq 1 ,0 \leq \ep \leq 1,
\lambda \geq 0 \right \}$, where
\bq
&&
{\cal \mu}_{\d , \ep , \lambda}(\rho_I )=
\no \no
&&
\sum_{i=1}^{k} Ad({\cal U}_{Bob}(\d , \ep))
Ad(W_i (\lambda))
Ad({\cal U}_{Alice}(\d , \ep)) \; \rho_I \;.
\nonumber
\eq
As previously the unitary operators implementing the gates of
Alice and Bob in the teleportation circuit are parameterized
by the imperfection parameters $\d , \ep$.
Then we observe that the dissipation operator on Bob's site
commutes with Alice unitary operation i.e., $[{\cal U}_{Alice},
W_i ]=0\;, i=\left\{1,\ldots, k\right\}$, so we have that
\bq
{\cal \mu}_{\d , \ep , \lambda}(\rho_I )=&&
\sum_{i=1}^{k} Ad({\cal U}_{tel}(\d , \ep))
Ad(W_i (\lambda)) \rho_I = 
\no \no
&&
{\cal U}_{tel}(\d , \ep))
s_\lambda (\rho_I )
{\cal U}_{tel}^\dagger (\d , \ep))\;.
\nonumber
\eq
At this point there are two ways to proceed. The first one
is based on the observation that the above dissipative
teleportation scheme is equivalent to the teleportation
scheme in which Alice and Bob share a mixed entangled state
and an enhancement of the quantum teleportation fidelity
is achieved by allowing either of them, to initially
perform a local
dissipative interaction with the environment\cite{horo}. Specifically
let  $V(.)=\sum_{i=1}^{k}V_i (.)V_i^\dagger $, then the
closeness of the initially shared bipartite state
$ {\bf 1}\otimes V (\rho )$, to the ideal maximally entangled state
$P_{EPR}=\proj{EPR}{EPR}$, is quantified by the
{\it fully entangled fraction}\cite{bene}, of the bipartite state,
$f=max_V Tr({\bf 1}\otimes V (\rho )P_{EPR})$. According to the
analysis of \cite{horo}, we search for such $V$ and $\rho$
that $f> 1/2$, so that the optimal fidelity of the teleportation
${\cal F}=\frac{2f+1}{3}$, exceeds the limit of the classical
communication viz. ${\cal F}_{cl}=\frac{2}{3}$.

Alternatively, we can simply proceed by assuming that
our dissipative holonomic teleportation has a lower
fidelity compared to the ideal teleportation scheme
and perform a first order perturbation say, of the
holonomic parameters $\d, \ep$
in order to estimate how close to one our fidelity can be.
Let us take the latter possibility and choose for definiteness
the phase damping mechanism described by the $k=2$ POVM, 
with $V_1 =diag(1 ,e^{-\lambda })$, and $V_2 =diag
(0,\sqrt{1-e^{-2\lambda}})$.
In terms of projectors $P_{ab}\equiv\proj{a}{b}$, the
initial state is written as $\rho_I^\Psi  =
\proj{\Psi}{\Psi}\otimes
P_{EPR}=\frac{1}{2}\sum_{i,j \in (0,1)}\a_i \overline{\a}_j P_{ij}
\otimes P_{EPR}$. Similarly if Alice measurement results in two
classical bits $(x,y)$,  then the final state of teleportation
is $\rho_F^\Psi =\proj{xy\Psi}{xy\Psi}=\frac{1}{2}
\sum_{k,l \in (0,1)}\a_k \overline{\a}_l P_{xx}\otimes
P_{yy}\otimes P_{kl}$. To estimate the quality of the dissipative
holonomic teleportation scheme against the standard ideal
teleportation we shall use the previously introduced fidelity factor
in its equivalent Hilbert-Schmidt or trace-norm form
\cite{nielchua}, which compares general mixed-state density operators and for 
our case reads:
\bq
&&
{\cal F}_{\d , \ep, \lambda}^{x,y}=min_{\ket{\Psi}}
Tr({\cal \mu}_{\d , \ep , \lambda}(\rho_I^\Psi )\rho_F^\Psi )\;.
\nonumber
\eq
After adding up all the different possibilities of $x$ and $y$ we obtain
for the fidelity up to the first order in $\ep$ and $\delta$, but to all orders in 
$\lambda$ the following expression
\bq
{\cal F}_{\d , \ep, \lambda}=&&{1 \over 2} \left( 1+e^{-\lambda} \right)-
\no \no
&&
\ep {3 \over 2} (\sqrt{2} -1)+ 
\ep (1-e^{-\lambda}) \left( {21 \over 32} \sqrt{7}-{51 \over 32} \right) -
\no \no
&&
\delta {1 \over 2 \sqrt{2} } +\delta (1 - e^{-\lambda}) {3 \over 16}
\sqrt{{ 3 \over 2}}\,\, .
\label{findiss}
\eq
The intriguing characteristic 
is that after allowing for dissipation to occur in the initial state by having non-zero 
values for $\lambda$ the coefficients of $\ep$ and $\delta$ become smaller, compared to the  dissipationless case of eq.(\ref{fid}). 
Analytically, let $\Delta{\cal F}_{\ep,\delta} = \frac{1}{2}-\ep
\frac{1}{32}(21\sqrt{7}-51)-\delta \frac{1}{4}\sqrt{\frac{3}{2}}$, then 
expression (\ref{findiss}) takes the form ${\cal F}_{\d , \ep, \lambda}=
{\cal F}_{\ep,\delta}-(1-e^{-\lambda})\Delta{\cal F}_{\ep,\delta}$.
Clearly the initial value ${\cal F}_{\ep,\delta}$, of the fidelity for zero dissipation
$\lambda = 0$, changes to the asymptotic non-zero value
${\cal F}_{\d , \ep, \infty}=
{\cal F}_{\ep,\delta}-\Delta{\cal F}_{\ep,\delta}$, for large 
dissipation $\lambda \longrightarrow \infty$. 
This signifies the fact that the fidelity of imperfect holonomic teleportation becomes  
resilient to some quantum dissipation that may occur during classical transmission  
of information.

{\em Conclusions.} In this work holonomic H and CN gates are employed for the
construction of the teleportation
circuit, which provides itself as an architectural unit for building in a
distributive and fault-tolerant way remote gates that effectively form a
universal set for quantum computation, by means of a procedure that requires only
prior searing of ancillary states between remote parties and quantum measurments. Although such a pivotal holonomic
circuit could be considered protected from quantum errors    
due to the geometrical functioning principle of its holonomic gates, possible 
errors coming from systematic geometrical imperfections of the construction of holonomies,
as well as from the presence of quantum noise during classical communication between
remote parties of the circuit, are almost unavoidable. The modeling and studying of these
types of errors showed that the holonomic teleportation circuit functioning under 
small imperfections of its gates is resilient to quantum noise during classical transmission.

Experimentally the theoretical construction presented here may be
materialized with ion traps. Such an implementation of
HQC, which uses the ion vibronic modes for the control manipulations presented
in \cite{PaChou} and enjoys exponential dumping in control errors, is currently
under investigation \cite{jianmpq}. Furthermore, the teleportation scheme with ion
traps presented in \cite{Solano} may be performed with holonomic gates
materializing practically the second part of our proposal and hence exhibiting
the resilience to both the control and quantum noise errors.

JP would like to thank Christof Zalka for useful conversations. 
JP acknowledges a TMR Network support under the contract no. ERBFMRXCT96-0087.

\end{multicols}


\begin{references}
\bibitem{compl} The tensor product structure, which is missing from the $\cp^n$ model \cite{PAZARA}
does not affect the results presented here on imperfections as well as the overall error of the circuits.
In the holonomic context the complexity may be faced successfully as in \cite{PaChou} and \cite{ams},
while the $\cp^n$ model is simple enough for presenting a complete study of control imperfections.
\bibitem{flux} The Abelianization of the theory by restricting on planes with commuting 
connection components eliminates the path ordering symbol and the treatment is as in 
the Abelian case.
\end{references}
\end{document}